\documentclass[journal=nalefd,manuscript=article]{achemso}
\usepackage{amsmath}
\usepackage{amssymb}

\usepackage[colorlinks,bookmarks=false,linkcolor=blue,urlcolor=blue]{hyperref}

\author{Ching-Hao Chang}
\affiliation{Institute for Theoretical Solid State Physics, IFW Dresden,
Helmholtzstr. 20, 01069 Dresden, Germany}
\email{c.h.chang@ifw-dresden.de}
\author{Carmine  Ortix}
\affiliation{Institute for Theoretical Physics, Center for Extreme Matter and Emergent Phenomena, Utrecht University, Princetonplein 5, 3584 CC, Utrecht, Netherlands}
\alsoaffiliation{Institute for Theoretical Solid State Physics, IFW Dresden,
Helmholtzstr. 20, 01069 Dresden, Germany}
\email{c.ortix@uu.nl}

\title{Theoretical prediction of a giant anisotropic magnetoresistance  in carbon nanoscrolls}


\begin{document}

\begin{tocentry}
\begin{center}
\includegraphics[width=.9\textwidth]{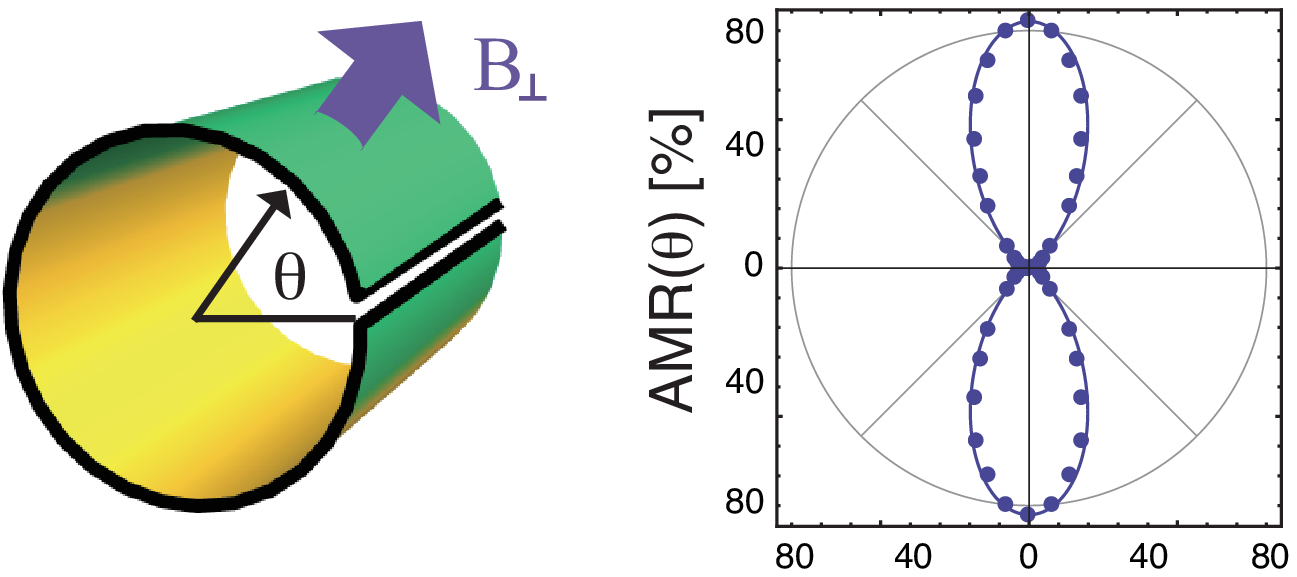}
\end{center}
\end{tocentry}

\begin{abstract}
Snake orbits are trajectories of charge carriers curving back and forth which form at an interface where either the magnetic field direction or the charge carrier type are inverted. In ballistic samples their presence is manifested in the appearance of magnetoconductance oscillations at small magnetic fields. Here we show that signatures of snake orbits can also be found in the opposite diffusive transport regime. We illustrate this by studying the classical magnetotransport properties of carbon tubular structures subject to relatively weak transversal magnetic fields where snake trajectories appear in close proximity to the zero radial field projections. In carbon nanoscrolls the formation of snake orbits leads to a strongly directional dependent positive magnetoresistance with an anisotropy up to $80 \%$. 
\end{abstract}

Carbon nanomaterials, such as carbon nanotubes (CNT) \cite{book-CNT} and graphene \cite{graphene}, continue to trigger a lot of attention due to their very unique structural and physical properties \cite{carbonreview}.
In recent years, another carbon nanomaterial, called carbon nanoscroll (CNS), has emerged \cite{nanoscroll}. 
It is a spirally wrapped graphite layer that, unlike a multiwalled carbon nanotube (MWCNT), is open at two edges and does not form a closed structure \cite{nanoscroll-structure}. 
Scroll whiskers were first reported by Bacon in 1960 \cite{nanoscroll-pressure}. 
A more recent chemical route \cite{nanoscroll} involves intercalation of graphite flakes with potassium followed by exfoliation with ethanol to form a dispersion of carbon sheets. Upon sonication, some of the exfoliated graphite sheets curl into scrolls. 
The disadvantage of this route is that CNSs are scrolled from an undefined number of graphene layers. In addition, the chemical process can potentially induce unexpected defects in the material, thereby lowering its quality. 
Controlled fabrication of high-quality CNSs has been instead achieved \cite{nanoscroll-nl} using isopropyl alcohol solutions to roll up high-quality monolayer graphene 
predefined on Si/SiO$_2$ substrates. 
CNSs formation obtained by rolling a graphene monolayer is dominated by two major energetic contributions \cite{nanoscroll-structure}: an elastic energy increase due to bending that decreases the CNS stability, and a free energy gain generated by the van der Waals interaction energy between the overlapping regions
of the scroll. Scroll stability requires a minimum layer overlap as well as a minimum inner diameter. 
Moreover, the inner diameter of the scroll has been predicted \cite{CNS-e-radius} to increase up to 2.5\% upon charge injection, 
which suggests that CNSs can be used as efficient electromechanical actuators at the nanometer scale \cite{CNS-transistor}. 
Finally, CNSs can also sustain high current densities, which facilitates their application as microcircuit interconnects\cite{nanoscroll-nl}.

The peculiar geometric structure of CNSs also yields unusual electronic \cite{CNS-metal}, optical \cite{pan05}, and transport properties in uniform electric and magnetic fields \cite{CNS-QT,scrollprl}. The natural presence of edge nanoscrolls in  graphene, for instance, has been predicted to be at the basis of the poor
quantization of the Hall conductance in suspended samples  \cite{scrollprl}. This is due to the fact that inside the scrolls, the electrons respond primarily to the normal component of the externally applied magnetic field \cite{ferrari-nt}, which oscillates in sign and largely averages out. 

In this work, we theoretically predict a strongly directional dependent magnetoresistance in CNSs subject to relatively weak transversal magnetic fields. The reason for the occurrence of this phenomenon is that the oscillation of the effective magnetic field felt by the electrons in a CNS leads to the formation of classical snake orbits, whose number changes with the direction of the externally applied magnetic field. As a result, we find a giant anisotropic magnetoresistance (AMR) with a magnitude of up to $80 \%$, a value comparable to the AMR observed in the quantum anomalous Hall phase of ferromagnetic topological insulator thin films\cite{amr-qahe}, and an order of magnitude larger than the bulk AMR of conventional ferromagnetic alloys \cite{amr}. 
This suggests a novel route towards miniaturized nanoscale devices exploiting the AMR effect for magnetic recording, for instance \cite{amr-book}.

To prove the assertions above, we first elucidate the effect of snake orbit formation by analyzing the magnetotransport properties of single-walled CNTs subject to transversal magnetic fields in the classical diffusive transport regime, where quantum-interference corrections to the conductivity are not expected to occur  \cite{gop15}. 
Since, as mentioned above, charge carriers respond to the radial normal projection of the transversal magnetic field we can switch from the native three-dimensional description to a two-dimensional (2D) one in which the CNT is modelled as a 2D channel subject to a periodic magnetic field of zero average $B(s) = B \cos (s/R_{\rm CNT})$, where $B$ is the strength of the externally applied transversal magnetic field, $R_{CNT}$ is the CNT radius, whereas $s= R_{CNT} \phi$ is the arclength in the tangential direction of the tube. Using Einstein relation and the classical linear response formula for the diffusion tensor, we can then write the conductivity tensor components as 
\begin{align}
\sigma_{i j}(B)=e^2 N \int_0^{\infty}\langle v_i (t)v_j (0)\rangle e^{-t/\tau}dt,
\label{eq:kubo}
\end{align}
where $i, j=\{s, y\}$, $y$ being the coordinate along the tube axis, $\tau$ is the relaxation time, $N$ the density of states, and $v_{i},v_{j}$ the carrier velocity components. The brackets $\langle \ldots \rangle$ denote an average over the available phase space, {\it i.e.} the velocity direction and the CNT azimuthal angle $\phi$. The velocity correlation function is obtained by solving the classical Newton equation of motion  $m^{\star} \, \dot{{\bf v}}= -e \,{\bf v}\times{\bf B}(s)$ with $m^{\star}$ the density-dependent dynamical mass \cite{bha16}. To obtain the conductivity tensor we have averaged the velocity correlation function with respect to 1600 sampling points, thereby reaching an accuracy within 1\%. We emphasize that the off-diagonal conductance $\sigma_{s y}$ identically vanishes due to the absence of an homogeneous magnetic field. 

\begin{figure}[h!]
\includegraphics[width=.6\columnwidth]{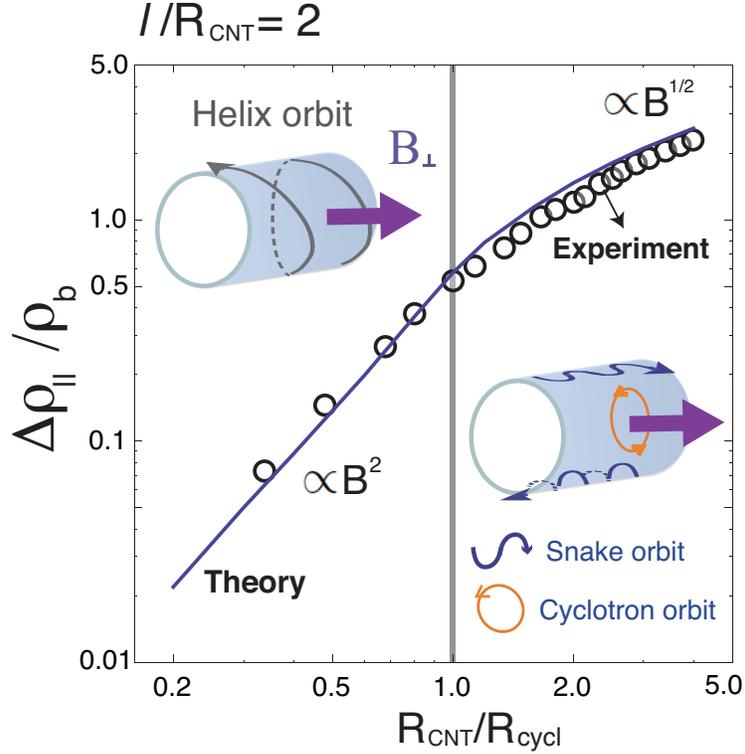}
\caption{(Color online). {\bf Classical magnetoresistance of a CNT.} Log-log plot of the MR as a function of the magnetic field strength $B$ measured by the ratio between the CNT radius $R_{\rm CNT}$ and the characteristic cyclotron radius $R_{\rm cycl}$. $\rho_{\rm b}$ is the longitudinal resistivity in the absence of externally applied magnetic fields. The circles are rescaled experimental results adapted from Ref. \cite{cnt-exp}.}
\label{fig:fig1}
\end{figure}

Fig.~\ref{fig:fig1} shows the ensuing behavior of the magnetoresistance $\Delta \rho_{\parallel} / \rho_{b} = \rho_{y y}(B) / \rho_{y y}(0) - 1$ as a function of the ratio between the CNT radius and the cyclotron diameter  $R_{cycl} = m^{\star} v_F / (e B)$ assuming a mean free path $l= v_F \tau$ twice as large as $R_{CNT}$. Since $R_{CNT} / R_{cycl} \propto B$,  Fig.~\ref{fig:fig1} provides us the magnetic field dependence of the magnetoresistance (MR). For low magnetic fields we find a positive MR, which grows quadratically with the magnetic field strength. A similar behavior has been experimentally found in nanotubules bundles \cite{son94} as well as in single MWCNTs \cite{cnt-exp}, and interpreted by considering a simple two-band model with an unequal number of electrons and holes contributing to the conductance \cite{not75}. Our calculations reveal instead that the positive MR arises in presence of a single type of charge carriers, and is due to the formation of characteristic helical orbits [see the inset of Fig.~\ref{fig:fig1} and the Supporting Information] wrapping the CNT during their motion. 
The quadratic growth of the magnetoresistance persists up to magnetic field strengths for which the cyclotron radius is comparable to the radius of the nanostructure.
It reaches a sizeable value of $\simeq  50 \%$  in Fig.~\ref{fig:fig1}. Even higher values of $\simeq 100 \%$ can be reached by considering a mean free path one order of magnitude larger than $R_{CNT}$ [see Supporting Information].  

In the regime where the cyclotron radius is smaller than  $R_{CNT}$ the behavior of the MR changes qualitatively. The MR indeed exhibits a much slower increase as testified by the slope change in Fig.~\ref{fig:fig1}. This is because the magnetic field is large enough to allow for the formation of usual cyclotron orbits 
localized in the regions of the CNT where the surface normal is parallel to the transversal magnetic field direction \cite{myprl,myreview},
which do not contribute to the MR. The latter is indeed entirely set by the contribution of the snake orbits appearing close to the boundaries where the normal component of the magnetic field switches its sign \cite{2degsnake}. As we explicitly prove in the Supporting Information, the snake state contribution to the MR is $\propto \sqrt{B}$ with a proportionality factor that is independent of the mean free path value in the regime $l \leq R_{CNT}$.
This sublinear power-law dependence of the MR for sufficiently large magnetic fields is in between the saturation \cite{saturMRreview,saturMR} and the linear growth \cite{graphenesheet,refereeP1,refereeP2} encountered in other material systems.

To verify the validity of our approach, we have compared our theoretical results with the MR measurements performed by Kasumov and collaborators  \cite{cnt-exp} on a $6$ nm outer radius isolated MWCNT, which show an inflexion point in the MR at an external moderate magnetic field of $\simeq 1.6$ T. 
From the condition that the inflexion point occurs when the CNT radius exactly matches the effective cyclotron radius, we obtain $m^{\star} v_F = 1.54\times 10^{-27}$ m $\cdot$ kg/s, which is compatible with a Fermi velocity \cite{CNT-fw,lia01} of the order of $10^5$ m/s and a cyclotron mass approximatively two order of magnitudes smaller than the mass of free carriers. 
By further taking into account a sizable magnetic-field independent resistivity, which we attribute to inter-wall and contact resistivities suppressing the MR by approximatively one order of magnitude, we find a perfect agreement in the behavior of the MR as a function of the magnetic-field strength [see Fig.~\ref{fig:fig1}]. Moreover, the value of the mean free path $l = 2 R_{CNT} \simeq 12$ nm is consistent with the experimental values reported in high-biased SWCNT \cite{nl-CNT-L}. 

\begin{figure}[tb]
\includegraphics[width=.9\columnwidth]{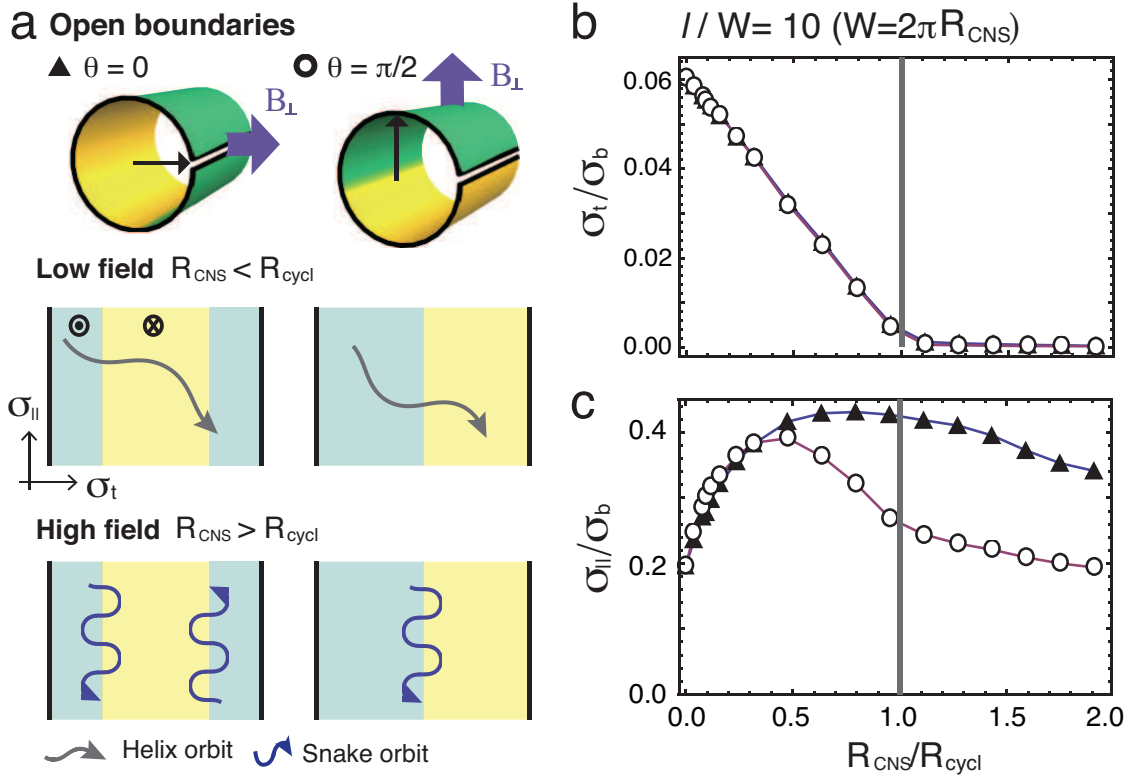}
\caption{(Color online). {\bf Electron orbits and magnetoconductivity of a one-winding CNS for different magnetic field directions.} {\bf a} The green and yellow regions indicate the portion of the CNS where the effective magnetic field felt by the electrons is positive and negative, respectively. The top panels schematically show the native three-dimensional description whereas the middle and bottom panels sketch the effective two-dimensional description with the characteristic electron trajectories in the weak and moderate field strength regime for different orientations.
{\bf b, c}, $\sigma_{\perp}$ ($\sigma_{\|}$) denotes the conductivity across (along) the tube axis, with $\sigma_{\rm b}$ the conductivity of a bulk 2D channel in the absence of magnetic fields. The triangles (circles) are the theoretical results for a one-winding CNS with mean free path $l/W=10$ subject to a field in the $\theta=0$ ($\theta=\pi/2$) direction.}
\label{fig:fig2}
\end{figure}  

Having established that our analysis in the classical diffusive transport regime correctly accounts for the behavior of the MR in CNTs up to moderate magnetic field strengths, we now move to analyze the magnetotransport properties of CNSs taking into account their peculiar geometric structure. The fact that the characteristic radii of the scrolls 
are comparable to the electronic mean free path -- they generally lie in the tens of nanometer scale -- implies that the electronic transport is in the quasi-ballistic regime \cite{be-review} where impurity scattering and boundary scattering at the inner and outer radius of the scroll are of equal importance. For the latter, we will assume fully diffusive scattering \cite{diffusiveB1,diffusiveB2} and set the specularity parameter to zero. 

In CNSs the adjacent graphene layers typically have incommensurate lattice structures. Therefore interlayer electron tunneling becomes negligible \cite{scrollprl} even though the separation between the overlapping regions is  about the same as in graphite, $0.34$ nm. This also implies that we can again switch to an equivalent two-dimensional description and analyze the magnetotransport properties of a 2D channel of total width $W$ subject to a periodic magnetic field with functional form $B=B \cos (s/R_{\rm CNS}+\theta)$. Here the angle $\theta$ indicates the direction of the transversal magnetic field with respect to the inner edge of the scroll with $s=0$. The outer edge of the CNS is defined by $s_{out}= 2 \pi w R_{CNS} $ with $w$ indicating the number of the CNS turns which is treated, for convenience, as a continuous variable \cite{ort10}. 
In the remainder we will restrict ourselves to a one-winding CNS with $w=1$. 
We emphasize that the presence of overlapping fringes in CNSs generally yields a noninteger value $w>1$. 
This, however, does not qualitatively change the main features of the electronic transport, which are entirely set by the presence, and not the precise location, of the open boundaries in a CNS. 

We can explicitly monitor the effect of boundary scattering on the magnetoconductivity by modifying Eq.~\ref{eq:kubo} as follows 
\begin{align}
\sigma_{ij}(B)= e ^2 N \int_0^{T}\langle v_i (t)v_j (0)\rangle e^{-t/\tau}dt,
\label{eq:kubo-T}
\end{align}
where $T$ is the transit time for a carrier with given initial position to reach the boundary \cite{li-size1,li-size2}.  
Fig.~\ref{fig:fig2}(b) shows the magnetic field dependence of the conductivity along the CNS azimuthal direction $\sigma_t = \sigma_{s s} $ measured in units of the conventional conductivity of a ``bulk" ($T \rightarrow \infty$) 2D channel in zero magnetic field. Here, we have set the mean free path $l$ to be one order of magnitude larger than the CNS width to assure the transport is well inside the quasi-ballistic regime. For zero magnetic field diffusive boundary scattering strongly suppresses the conductivity along the CNS width. A finite magnetic field leads to a further decrease of the conductivity, independent on the direction of the transversal magnetic field. The behavior of the conductivity component along the CNS axis $\sigma_{\parallel} = \sigma_{y y}$ is instead entirely different [see Fig.~\ref{fig:fig2}(c)]. In the weak-field regime $R_{CNS} << R_{cycl}$ we find an enhancement of the conductivity due to magnetic reduction of backscattering \cite{be-review}. This enhancement of the conductivity is followed by an ultimate suppression due to the formation of snake orbits, which, as discussed above, yield a positive MR.  Moreover, we find the conductivity $\sigma_{\parallel}$ to strongly depend on the magnetic field direction. This is because 
for a magnetic field oriented along the edge axis ($\theta=0$) there are two regions where its normal component switches sign [c.f. Fig.~\ref{fig:fig2}(a)], contrary to the case of a magnetic field oriented perpendicularly to the edge axis ($\theta=\pi/2$) in which case the magnetic field switch is encountered only along one line of the scroll. The ensuing proliferation of snake orbits for $\theta=0$ then leads to a much slower suppression of the conductivity since their contribution $ \propto 1/\sqrt{B}$ instead of the usual $1/B^2$ contribution of cyclotron orbits.

\begin{figure}[tb]
\includegraphics[width=.5\columnwidth]{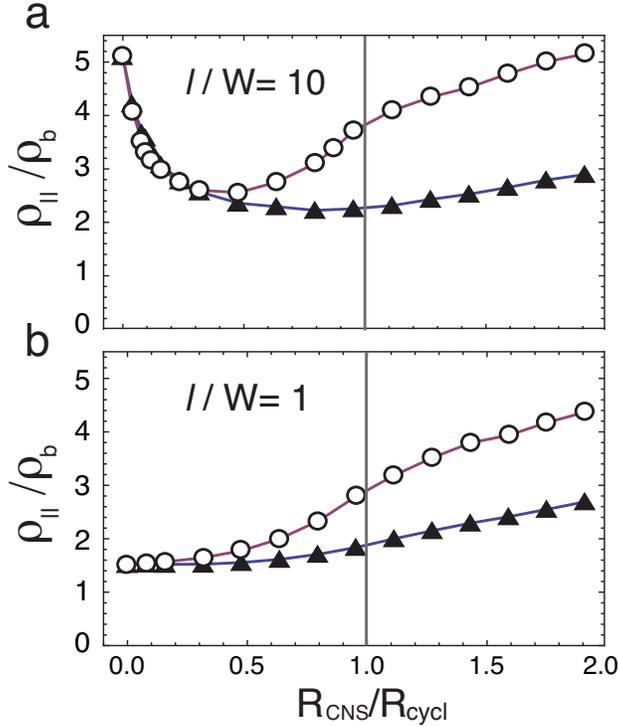}
\caption{(Color online).  {\bf Classical magnetoresistance of a one-winding CNS.} MR as a function of the magnetic field strength $B$ measured by the ratio between the CNS radius $R_{\rm CNS}$ and the characteristic cyclotron radius $R_{\rm cycl}$. $\rho_{\rm b}$ is the longitudinal resistivity of a bulk 2D channel in the absence of magnetic fields. The triangles are the result for a magnetic field direction $\theta=0$ while the circles are for $\theta=\pi/2$. 
The mean free path has been set to $l= 10 W$ and $l=W$ in (a) and (b) respectively.}
\label{fig:fig3}
\end{figure}  

The knowledge of the magnetoconductivity tensor components allow us to obtain the behavior of the magnetoresistance $\rho_{\parallel}$. 
For $ l / W \lesssim 10$, the zero-field resistivity is well described by the well-known formula \cite{be-review} $\rho_{\|} = \rho_b \left(1+\frac{4}{3\pi} \frac{l}{W} \right) $ accounting for boundary scattering effects on the resistivity. In the weak-field regime a negative MR due to magnetic suppression of backscattering is explicitly manifest only when the mean free path largely exceeds the width of the CNS [c.f. Fig.~\ref{fig:fig3}], which is in perfect analogy with the situation encountered in a 2D channel subject to an homogeneous perpendicular magnetic field \cite{DL-size}. In the intermediate field regime, the MR behavior strongly resembles the MR in the absence of boundary scattering [c.f. Fig.~\ref{fig:fig1}] but acquires a strong directional dependence independent of the ratio $l/W$. As long as the boundary scattering is completely diffusive, the directional dependence comes entirely from the aforementioned proliferation of snake orbits for $\theta=0$ independent of the relative importance between internal impurity and boundary scattering. 

\begin{figure}[tb]
\includegraphics[width=.6\columnwidth]{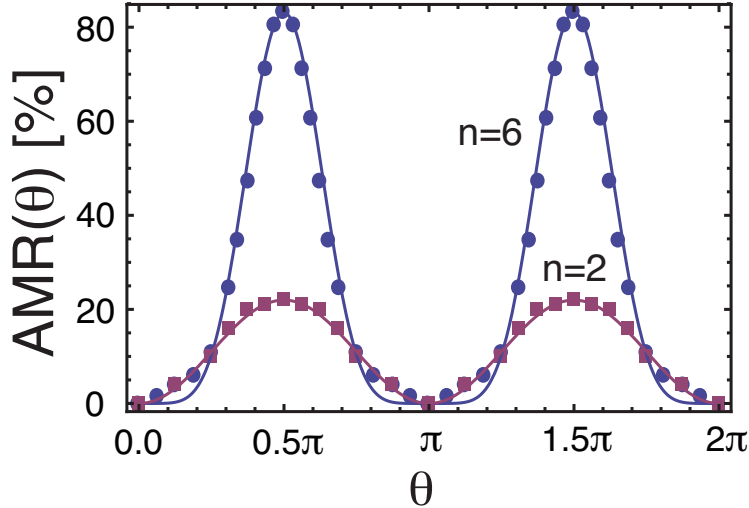}
\caption{(Color online). {\bf Anisotropic magnetoresistance of a one-winding CNS.} Angular dependence of  the AMR $\rho_{\parallel}(\theta) / \rho_{\parallel}(0) - 1 $ for $R_{\rm CNS}/R_{\rm cycl}=0.64$ (squares) and $R_{\rm CNS}/R_{\rm cycl}=3$ (disks). The ratio between the mean free path and total width of the CNS has been set to $l/W = 10$.}
\label{fig:fig4}
\end{figure}  

Fig.~\ref{fig:fig4} shows the dependence of the magnetoresistance as a function of the external magnetic field direction. We find that the angular dependence of the magnetoresistance takes the functional form 
\begin{align}
\rho(\theta)=\rho_{\parallel}(0)+ \delta \rho_{\parallel} \sin^n\theta, 
\label{eq:amr}
\end{align}
with $\delta \rho_{\parallel} \equiv \rho_{\parallel} (\theta=\pi/2) - \rho_{\parallel} (0)$ quantifying the magnitude of the anisotropic magnetoresistance (AMR) effect. 
In the weak field regime $R_{CNS}<R_{cycl}$ the angular dependence of the MR can be accurately described by the functional form of the AMR effect in bulk materials \cite{amr} given by $n=2$ in Eq.~\ref{eq:amr}. Its magnitude is proportional to the strength of the externally applied magnetic field, and can be further increased under charge injection due to the CNS diameter increase \cite{CNS-e-radius}.
 For sufficiently large magnetic field strengths, however, the increase of the resistance at $\theta=\pi/2$ becomes much more rapid with the angular dependence that can be described by Eq.~\ref{eq:amr} with $n=6$. In this regime and independent of the $l/W$ ratio, the AMR reaches a giant value $\simeq 80\%$, which is comparable to the AMR magnitude observed  in the quantum anomalous Hall phase of ferromagnetic topological insulator thin films\cite{amr-qahe}.

To wrap up, we have predicted, using a simple model of classical diffusion, a giant anisotropic magnetoresistance in carbon nanoscrolls subject to externally applied transversal magnetic fields. This phenomenon is entirely due to the formation of snake orbits yielding a positive magnetoresistance, whose number changes with the direction of the field. For moderate magnetic fields for which the effective cyclotron radius is smaller than the characteristic radius of the CNS we find an extremely large AMR with a magnitude up to $80 \%$. 
In the ballistic regime, snake states formed in ultraclean graphene $p-n$ junctions \cite{pnsnake-prl} have been shown to lead to conductance oscillations \cite{pnsnake-nc1,pnsnake-nc2}. A similar phenomenon has been predicted to occur in low-density semiconducting core-shell nanowires subject to transversal magnetic fields\cite{snakenl}. Our calculations proves on solid grounds that the formation of snake orbits strongly impacts the classical diffusive transport properties as well. 
Therefore, our prediction can be tested not only in CNS but also in semiconducting curved nanostructures manufactured with the rolled-up nanotechnology \cite{runt-nature}. 

\subsection*{Supporting Information}
Classical electron orbits in tubular nanostructures subject to transversal magnetic fields, snake orbits contribution to the magnetoresistance, details of the angular dependence of the magnetoresistance. This material is available free of charge via the Internet at http://pubs.acs.org. 

\subsection*{Notes}
The authors declare no competing financial interests.

\begin{acknowledgement}
We acknowledge the financial support of the Future and Emerging Technologies (FET) programme within
the Seventh Framework Programme for Research of the European Commission 
under FET-Open grant number: 618083 (CNTQC).  C.O. acknowledges support from the Deutsche Forschungsgemeinschaft (Grant No. OR 404/1-1), and from a VIDI grant (Project 680-47-543) financed by the Netherlands Organization for Scientific Research (NWO). C-H. C. thanks Kun Peng Dou for helpful discussions.
\end{acknowledgement}

\bibliography{runt}

\providecommand{\latin}[1]{#1}
\providecommand*\mcitethebibliography{\thebibliography}
\csname @ifundefined\endcsname{endmcitethebibliography}
  {\let\endmcitethebibliography\endthebibliography}{}

\end{document}